\documentstyle[preprint,aps,epsf]{revtex}
\begin{document}
%\pagestyle{empty}
%\twocolumn[\hsize\textwidth\columnwidth
%\hsize\csname@twocolumnfalse\endcsname
\preprint{\parbox[b]{3.3 cm} { hep-ph$/$9712366\\
aps1997dec13$\_$003\\ 
IP-ASTP-05-97 } }
\draft
\vfill
\title{Twist-3 and Quark Mass Contributions \\
to the Polarized Nucleon Structure Function $g_2(x,Q^2)$}
\draft
\vfill
\author{Kwei-Chou Yang \footnote{Email address: 
{\tt kcyang@phys.sinica.edu.tw}} and Hoi-Lai Yu \footnote{Email address:
{\tt PHHLYU@ccvax.sinica.edu.tw}}}
\address{Institute of Physics, Academia Sinica, Taipei, Taiwan 115, R.O.C.}
\maketitle
%\date{June 1997}
%%%%%%%%%%%%%%%%%%%%%%%%%%%%%%%%%%%%%%%%%%%%%%%%%%%%%%%%%%%%%%%%%%%%%%%%%%
\begin{abstract}

Quark mass effects are clarified in the parton model approach to the
transversely polarized nucleon structure function. The special propagator
technique is employed to obtain manifestly gauge invariant results and extract
the buried short-distance contributions inside the soft part after momentum
factorization  in the collinear expansion approach. A generalized massive 
special propagator for a massive quark is constructed. We identify the 
corresponding matrix elements of the transversely polarized structure function 
in deep inelastic scatterings by the massive special propagator technique.
\end{abstract}


%\section{INTRODUCTION}
     The naive parton model that built upon massless free partons had
proved itself tremendously successful towards the understanding of spin 
averaged high energy processes. However, a lot of interesting physics and 
even possible spin dependent new physics are washed out during the 
averaging procedure.  Following the advances of experimental techniques 
and facilities, both polarized probe and target become more popular. 
The EMC data~\cite{emc} on longitudinally polarized deeply inelastic 
scattering (DIS) experiments had already provided us lots of surprises 
and insights into the nuclear structure in the past decade. Spin physics 
has therefore become one of the most fascinating subjects towards the 
understanding of the quark and gluon dynamics inside hadrons. High 
precision data along with state-of-the-art higher order perturbative 
QCD (pQCD) computation enable us to test the standard model to high 
accuracy.  However, in DIS, where most spin data exist, much work on the 
subject has concentrated on the leading twist contributions which measure 
the helicity of the quark constituents. Recently, results  in the DIS 
transversely polarized structure function $g_T(x,Q^2)$ were 
reported and have shown a non-negligible contribution~\cite{smc}.
$g_T$ contains a chiral even part which measures the quark transverse
spin asymmetry and a  chiral odd part which is described by  the quark
transversity distribution~\cite{rat,jj} in the nucleon.
Extensive study is expected to be performed at DESY, CERN and SLAC.

     Quark transverse spin is famous for its conceptual difficulties and
confusions in the literature~\cite{rat,cor,art,jaffe}.  As emphasized in 
Ref.~\cite{art}, quark transverse spin is a fundamental degree of freedom, 
and the transversity parton distribution which measures the quark
helicity-flip in the helicity basis is well-defined even for massless partons.
To see its partonic probabilistic interpretation one has to go to
transverse spin eigenstate where transversity becomes diagonal. It simply
measures the difference of oppositely transverse polarized quarks 
inside the nucleon.  
Using an anti-quark probe as in the Drell-Yan process, the transversity 
is a leading twist effect and is a naturally large quantity.
It is important to note that helicity and chirality are identical for ``good"
light cone component of the Dirac field. Since in DIS,
the virtual photon is a chirally invariant probe, upon neglecting the small
quark mass, the quark chirality becomes a good quantum number, and thus
renders the quark transversity, no matter how large, invisible
by the virtual photon probe in DIS. It is this different chiral transformation
property between the ``probe" and ``parton distribution" that causes much 
confusion in the literature.  The situation becomes even worse
in the operator product expansion for DIS, where an unambiguous
separation of the probe's and operators' chiral transformation properties is
more difficult.  To summarize, the quark transversity
is a measurement of chiral symmetry breaking effects and mixes with 
other complicated high twist (twist-3) transverse spin asymmetry 
contributions in DIS.

   To measure the transverse quark spin in DIS requires the 
inclusion of quark masses and hence is of high twist in nature. However, it 
is well known that the quark mass term in the final state does
not respect the electromagnetic gauge invariance. The authors in 
Refs.~\cite{et,ael} have shown that it is necessary to include the ``twist-3" 
gluon term (i.e. the transverse momentum) and use the equation of motion to 
achieve a gauge invariant final result. However, the mixing of multiparton 
contributions makes the parton picture very unclear.
It is therefore of great importance to identify the twist-3
contributions to transversely polarized DIS within a generalized massive
parton model in a consistent and systematic way. A well-defined collinear 
factorization algorithm to identify the non-leading twist matrix elements 
that involve the incorporation of parton transverse momentum had already 
existed in the literature~\cite{efp,qiu}.  All these works have neglected 
the mass of the parton. To investigate the parton mass effects, the authors 
in Ref.~\cite{bt} introduce the spurion which couples only to the massless 
parton. This procedure leads to correct answers but it loses 
the trace of the symmetry breaking effects of the above-mentioned chirality 
selection rule in DIS. In view of the conceptual importance of the quark 
mass at hand, we feel that it would be of more transparency both conceptually 
and technically to deal directly with a massive parton. The introduction of 
the extra quark mass $m_q$ will not cause any inconsistency to the 
originally single scale collinear factorization algorithm, as long as  
it is much less than the factorization scale in the problem.

  In the following, we shall follow Qiu~\cite{qiu} and introduce a generalized
special propagator for massive quarks. The advantage of Qiu's special
propagator method is to completely separate the hard part between different 
orders in 1/Q (twist) in a manifestly electromagnetic gauge invariant way,
which is crucial for the problem at hand. 
The idea is to extract the hidden hard part from an apparently soft part 
after spinor and Lorentz index factorizations. Contrary to the conventional 
claim that the high twist matrix elements are lack of a simple
parton model interpretation due to mixing between matrix elements of
various numbers of partons, Qiu's approach will pick up only a fixed number of
partons in each particular twist and therefore makes a good simple
parton-model interpretation of the matrix element possible.

     The antisymmetric part $W_{\mu\nu}^A$ of the hadronic tensor $W_{\mu\nu}$
in DIS which describes the QCD spin physics is
\begin{equation}
\frac{W^A_{\mu\nu}}{2M_N}=\frac{1}{P\cdot q}i\epsilon_{\mu\nu\alpha\beta}
q^{\alpha} \left[
S^\beta g_1(x,Q^2) + \left( S^\beta - \frac{S\cdot  q}{P\cdot  q}
P^\beta \right) g_2(x,Q^2) \right],
\end{equation}
where $P$, $S$, $M_N$ and $q$ are the momentum, spin vector, mass of the
nucleon and momentum of the virtual photon probe, respectively.
We introduce two light-like vectors $n^{\mu}={\delta}^{{\mu}-}$ and
${\bar n^\mu}={{\delta}^{\mu+}}$ for our coordinate. In the frame in
which the proton with momentum $P$ is moving in the  z-direction,
one can parameterize $P^\mu$, $q^\mu$, and the proton spin vector $S^\mu$ as
\begin{eqnarray}
P^\mu&&=p \bar n^\mu + {M_N^2\over 2p}n^\mu,\nonumber\\
q^\mu&&=-x_B(1-{x_B^2 M_N^2\over Q^2}) p \bar n^\mu +
{Q^2\over 2x_B p}(1+{x_B^2 M_N^2\over Q^2})n^\mu\nonumber\\
&&\equiv -\tilde xp\bar n^\mu+{Q^2\over 2 \tilde xp}n^\mu,\nonumber\\
S^\mu&&=(S\cdot n)\Bigl(\bar n^\mu -{M_N^2\over 2p^2}n^\mu\Bigr)
+S^\mu_\bot,
\end{eqnarray}
where $x_B={Q^2\over2P\cdot q}$, $S^2=-1$, and we have assumed
${{M_N^2}\over Q^2}\ll 1$. In this frame, the parton momentum $k^\mu$ 
can be decomposed as
\begin{eqnarray}
k^\mu=\widehat k^\mu + {k^2-m_q^2\over 2k\cdot n}n^\mu,
\end{eqnarray}
where
\begin{eqnarray}
\widehat k^\mu\equiv(k\cdot n)\bar n^\mu +
{k^2_\bot +m_q^2\over 2k\cdot n}n^\mu
+ k_\bot^\mu,
\end{eqnarray}
is the on-shell part, satisfying
$$
\widehat k^2=m_q^2,
$$
with $m_q$ being the parton-quark mass.

  With the above momentum parametrization the quark propagator with
momentum $k$ can be decomposed as
\begin{eqnarray}
{i(\not k+m_q)\over k^2-m_q^2}=
{i(\not\widehat k+m_q)\over k^2-m_q^2}
+{i\not\! n\over 2k\cdot n},
\end{eqnarray}
where the $i\not n\over 2k\cdot n$ term is known as the special propagator in
Ref.~\cite{qiu}. Note that the form of the special propagator is the same
as that in the massless parton case. This is consistent with the short 
distance property of the special propagator.  It was pointed out by Qiu that 
this special propagator offers no spatial separation along light-cone.
Therefore, when the soft part, say, the ``naive" twist-2 matrix element
in our DIS case,
\begin{eqnarray}
\widehat T(k)=\int dz e^{ikz} \langle PS|\bar\psi(0) \psi(z)|PS\rangle
\label{eq:t}
\end{eqnarray}
is contracted with $\not\!\!\bar n$ after the EFP~\cite{efp} collinear factorization
procedure, it will actually contain a hidden short distance contribution even 
in the zero transverse momentum ${\bf k}_\perp$ limit. In particular, after 
extracting an extra quark-gluon vertex and a special propagator into the 
hard part, the new soft part will contain one more gluon, and becomes a 
twist-3 matrix element,
\begin{eqnarray}
\widehat T^\alpha(k,k')=\int dz_1 dz e^{i(k_1-k)z_1} e^{ikz}
\langle PS|\bar \psi(0)(-g_s T^a A_a^\alpha(z_1)) \psi(z)|PS\rangle.
\label{eq:ta}
\end{eqnarray}
Without removing this hidden short distance contribution, one will suffer
from the ambiguous mixing of soft parts between different ``twists". 
For example, one will have to use the equation of motion to link up 
$\widehat T$ and $\widehat T^\alpha$, which will invalidate the naive parton 
model interpretation of these soft matrix elements. Another important
feature of this special propagator is to extract also the ${\bf k}_\perp$
contributions in $\widehat T$ (which is of higher twist by definition).
After combining with the gluon field $A^\alpha$ in $\widehat T^\alpha$,
a color gauge invariant covariant derivative can be
achieved. To see this, 
we contract the loop parton propagator with $\not\!\bar n$ (contraction with
$\not\! n$ leads to leading twist results), which gives
\begin{eqnarray}
{i(\not\widehat k+m_q)\over k^2-m_q^2}\not\! \bar n=&&
{i(\not k+m_q)\over k^2-m_q^2}{\not k-m_q\over k^2-m_q^2}
(\not\widehat k+m_q) \not\! \bar n\nonumber\\
=&&{i(\not k+m_q)\over k^2-m_q^2}[(k-xp\not\! \bar n)^\alpha (i\gamma_\alpha)-
im_q]{i\not\! n\over 2k\cdot n} \not\!\bar n,
\label{eq:sp}
\end{eqnarray}
where we have used $\widehat k^2=m_q^2$ and the collinear expansion
$k\cdot n=\widehat k\cdot n =2xP\cdot n$. It is clear from Eq.~(\ref{eq:sp})
that the hidden effective vertex $i(\not k-xp\not\!\!\bar n-m_q)$ should be moved
into the hard part and classified as a high-twist contribution due to the
presence of the special propagator $i\not n\over 2k\cdot n$. Before proceeding,
we would like to point out that the introduction of the quark mass $m_q$
does not alter the procedures of having the special propagator to extract
the hidden short distance contribution from the apparent soft part after
the EFR collinear factorization. This should be obvious since the special
propagator is of short distance in nature, and should not be affected by
the presence of the quark mass.

     We are now ready to identify the twist-three matrix elements response for
the transverse polarization in DIS. We shall begin with the virtual-photon
hadron forward Compton scattering,
which is
\begin{eqnarray}
T^{\mu\nu}=\int \frac{d^4 k}{(2\pi)^4}[\widehat S^{\mu\nu}(k)\widehat T(k)]+
\int \frac{d^4 k_1}{(2\pi)^4}\frac{d^4 k}{(2\pi)^4}
[\widehat S^{\mu\nu}_\alpha(k_1,k)\widehat T^\alpha (k_1,k)]+\cdots,
\label{eq:tuv}
\end{eqnarray}
where Disc($T^{\mu\nu}$)=2$\pi iW^{\mu\nu}$, $\widehat T(k)$ and 
$\widehat T^\alpha(k_1,k)$ are the same as in Eqs.~(\ref{eq:t}) and 
(\ref{eq:ta}) respectively. To pick up the twist-3 contributions, 
we first collinearly expand the ``hard" part $\widehat S^{\mu\nu}$ and 
$\widehat S^{\mu\nu}_\alpha$ up to the relevant order of interest:
\begin{eqnarray}
\widehat S^{\mu\nu}(k)&&=\widehat S^{\mu\nu}(xp\bar n)+
\frac{\partial \widehat S^{\mu\nu}}{\partial k^\alpha} |_{k=xp\bar n}
(k-xp\bar n)^\alpha
+\cdots,\\
\widehat S^{\mu\nu}_\alpha(k_1,k)&&=\widehat S^{\mu\nu}_\alpha
(x_1p\bar n,xp\bar n)+
\frac{\partial \widehat S^{\mu\nu}_\alpha}{\partial k^\beta_i} |_{k_i=x_ip\bar n}
(k_i-x_ip\bar n)^\beta
+\cdots.
\label{eq:sa}
\end{eqnarray}
Before proceeding, some remarks are in order. Naively, one would expect that
the second term in Eq.~(\ref{eq:sa}) belongs to the twist-4 contribution
and should be dropped in the twist-3 discussion. However, a careful
investigation indicates that this term will actually give rise to a
${\bf k}_{\perp\beta}A_\alpha$ contribution after the Lorentz index 
separation. If it is true, then one would obtain a twist-3 matrix element that 
contain quark fields and a single gluonic field strength as in the single
transverse spin asymmetry case\cite{qs}. Further studies reveal that it is
not the case in DIS. The soft gluon pole cancels each other between mirror 
diagrams (with respect to the Cutkosky cut) in the hard part.
So, we shall just drop the second term in Eq. (\ref{eq:sa}) in what follows. 
Using the Ward identity,
\begin{eqnarray}
\frac{\partial}{\partial k^\alpha}
\frac{i}{\not k+\not q-m_q}=
\frac{i}{\not k+\not q-m_q}(i\gamma_\alpha)
\frac{i}{\not k+\not q-m_q},
\end{eqnarray}
we obtain
\begin{eqnarray}
\frac{\partial \widehat S^{\mu\nu}}{\partial k^\alpha} |_{k=xp\bar n}
=\widehat S^{\mu\nu}_\alpha(xp\bar n, xp\bar n).
\end{eqnarray}
Inserting the identity $1=\int dx \delta(x-{k\cdot n\over P\cdot n})$ into
Eq.~(\ref{eq:tuv}) and with the help of the identity
$$\int {d^4 k\over (2\pi)^4} e^{ikz} \delta(x-{k\cdot n\over P\cdot n})
=\int {d\lambda\over 2\pi} e^{i\lambda x}\delta^{(4)}
(z-{\lambda\over P\cdot n}n),$$
we can integrate out the uninteresting $k_-$ and ${\bf k}_\perp$ components and
arrive at
\begin{eqnarray}
T^{\mu\nu}=\int dx_1 dx {\rm Tr}[\widehat S^{\mu\nu}(x_1p\bar n,xp\bar n)
T(x_1,x)]+
\int dx_1 dx {\rm Tr}[\widehat S^{\mu\nu}_\alpha(x_1p\bar n,xp \bar n)
T^{\alpha}(x_1,x)],
\label{eq:Tuv}
\end{eqnarray}
where
\begin{eqnarray}
T_{ij}(x_1,x)&&=\int {d\eta\over 2\pi} {d\lambda\over 2\pi}
 e^{i\eta(x-x_1)} e^{i\lambda x_1}
\langle PS|\bar \psi_j(0) \psi_i({\lambda n\over P\cdot n})
|PS\rangle,\nonumber\\
T_{ij}^{\alpha}(x_1,x)&&=\int {d\eta\over 2\pi}
{d\lambda\over 2\pi}
 e^{i\eta(x-x_1)} e^{i\lambda x_1}
\langle PS|\bar \psi_j(0)D^{\alpha'}({\eta n\over P\cdot n})
\psi_i({\lambda n\over P\cdot n})
|PS\rangle\omega^\alpha_{\ \alpha'},
\label{eq:Ta}
\end{eqnarray}
with $D^{\alpha'}=i\partial^{\alpha'}-g_s T^a A_a^{\alpha'}$,
$\omega^\alpha_{\ \alpha'}=g^\alpha_{\ \alpha'}-\bar n^\alpha n_{\alpha'}$
being the covariant derivative and projection operator, respectively.
In the above, we have suppressed the quark flavor index for simplicity.
Note that we have employed the light-cone gauge $n\cdot A=0$ and therefore
$\omega^\alpha_{\alpha'} A^{\alpha'}[\eta/(p\cdot n)]=
A^{\alpha}[\eta/(p\cdot n)]$ to arrive at Eq.~(\ref{eq:Ta}).
One can easily show that a path order link-operator should be inserted in
Eq.~(\ref{eq:Ta}) if non-light-cone gauges are employed.

   It is important to note that in this special propagator formalism, the
hard part is the sum of the ``conventional" and the special propagator 
contributions. In terms of Feynman diagrams and obvious notations, we have

%\begin{figure}
%\vspace{2.5cm}
%\hspace{3.1cm}
\epsfbox{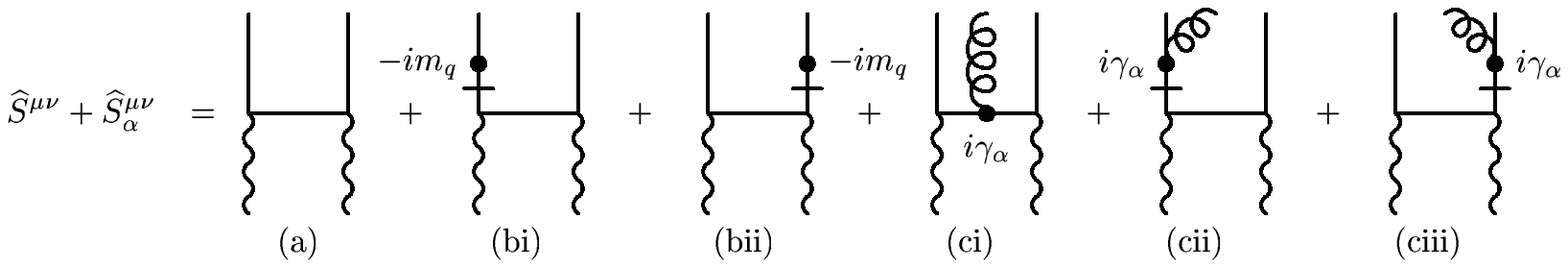}
\vspace{2.8cm}
%\caption{\label{fig:trace1}
%\end{figure}
\noindent where
\begin{equation}
\begin{picture}(50,20)(0,5)
\put(5,5){\line(1,0){30}}
\put(20,0){\line(0,1){10}}
\put(5,12){\vector(1,0){10}}
\put(3,5){\makebox(0,0)[r]{$\scriptstyle j$}}
\put(37,5){\makebox(0,0)[l]{$\scriptstyle i$}}
\put(5,15){\makebox(0,0)[bl]{$\scriptstyle k$}}
\end{picture}
=\frac{i\mbox{$\not\! n\,$}_{ij}}{2k\cdot n}
\end{equation}
is the special propagator in the above diagram. For simplicity, we have also
omitted the cross diagrams for the virtual photon which corresponds to the 
antiquark contributions.  After this lengthy discussion on the gauge invariant 
collinear expansion, we are ready to perform factorization in spinor 
indices, which is basically a Fierz transformation.  For the two-parton matrix 
element, we can expand the Dirac matrix in terms of the 16 independent 
orthogonal bases. The relevant terms for our purposes are
\begin{eqnarray}
&&\int\frac{d\lambda}{2\pi}e^{i\lambda x}\langle P,S |\bar\psi_j(0)
\psi_i(\frac{\lambda n}{Pn})|P,S\rangle\nonumber\\
&&=-M_N (S\cdot n)  g_1(x) (\not\! \bar n \gamma_5)_{ij}+
\frac{i}{2} (P\cdot n) h_1(x)\bar n^\alpha S^\beta_\bot
(\sigma_{\alpha\beta}\gamma_5)_{ij}+\dots,
\label{eq:2p}
\end{eqnarray}
where
\begin{eqnarray}
&&\int \frac{d\lambda}{8\pi} e^{i\lambda x} \langle PS|\bar \psi(0)
\not\! n \gamma_5 \psi\Bigl[\frac{\lambda n}{P\cdot n}\Bigl]|PS \rangle
=M_N (S\cdot n)  g_1(x),\nonumber\\
&&\int \frac{d\lambda}{4\pi} e^{i\lambda x} \langle PS|\bar \psi(0)
\not\! n  d^{\beta\sigma} \gamma_\sigma \gamma_5
\psi\Bigl[\frac{\lambda n}{P\cdot n}\Bigl]|PS \rangle
=(P\cdot n) d^{\beta\sigma}S_\sigma h_1(x).
\end{eqnarray}
Likewise, the expansion for the three-parton matrix elements relevant for
discussion is:
\begin{eqnarray}
&&\int \frac{d\lambda}{2\pi}\frac{d\eta}{2\pi} e^{i\lambda x_1}
e^{i\eta(x_2-x_1)} \langle PS|\bar \psi_j(0)
D^\alpha\Bigl[\frac{\eta n}{P\cdot n}\Bigr]\psi_i\Bigl[\frac{\lambda n}
{P\cdot n}\Bigr] |PS\rangle\nonumber\\
&&={i\over 2}(P\cdot n)M_N G(x_1,x_2)S_{\bot\delta}
\epsilon^{\alpha\delta}_{\bot} \not\!\bar n_{ij}-
{1\over 2} (P\cdot n) M_N \widetilde G(x_1,x_2)S^\alpha_\bot
(\not\!\bar n\gamma_5)_{ij}+ \dots,
\label{eq:3p}
\end{eqnarray}
where we have used
$S^\beta_\bot=-d^{\beta\lambda}S_\lambda$, $\epsilon^{\beta\delta}_\bot=
\epsilon^{\beta\delta\lambda\sigma}\bar n_\lambda n_\sigma$, and
$d^{\beta \lambda}$=diag(0, 1, 1, 0) is the projection operator in the transverse
direction.  Inverting Eq.~(\ref{eq:3p}), $\widetilde G$ and $G$  can be
 obtained as follows: 
\begin{eqnarray}
&&\int \frac{d\lambda}{2\pi} \frac{d\eta}{2\pi} e^{i\lambda x_1}
e^{i\eta(x_2-x_1)} \langle PS|\bar \psi(0)\not\! n \gamma_5
d^{\beta\alpha}
D_\alpha\Bigl[\frac{\eta n}{P\cdot n}\Bigr]\psi\Bigl[\frac{\lambda n}
{P\cdot n}\Bigr] |PS\rangle\nonumber\\
&&=-2 S^\beta_\bot (P\cdot n) M_N \tilde G(x_1,x_2),\nonumber\\
&&\int \frac{d\lambda}{2\pi}\frac{d\eta}{2\pi} e^{i\lambda x_1}
e^{i\eta (x_2-x_1)}
\langle PS|\bar\psi(0) \not\! n\epsilon^{\beta\delta}_\bot
D_\beta\Bigl[\frac{\eta n}{P\cdot n}\Bigr]
\psi\Bigl[\frac{\lambda n}{P\cdot n}\Bigr] |PS \rangle\nonumber\\
&&=-2iS_\bot^\delta (P\cdot n) M_N G(x_1,x_2).
\end{eqnarray}
We now perform the Lorentz index factorization, which can be achieved by
decomposing the metric tensor
\begin{eqnarray}
g_{\alpha\beta}=\bar n_\alpha n_\beta +\bar n_\beta n_\alpha -d_{\alpha\beta}.
\end{eqnarray}
Substituting Eqs.~(\ref{eq:2p}),~(\ref{eq:3p}) into Eq.~(\ref{eq:Tuv}),
we finally arrive at the factorization
formula for ${\rm\ Im}\ T^{\mu\nu}$ in DIS
\begin{eqnarray}
{\rm Im\ }T^{\mu\nu}=&&\int dx_1 dx
\Bigl [ \delta(x-x_1) g_1(x)\sigma_a^{\mu\nu}(x)+\delta(x-x_1)
h_1(x) \sigma_b^{\mu\nu}\nonumber\\
&&+G(x_1,x)\sigma_{c_1}^{\mu\nu}(x_1,x)
+\tilde G(x_1,x)\sigma_{c_2}^{\mu\nu}(x_1,x)
\Bigr],
\end{eqnarray}
and
\begin{eqnarray}
\sigma_a^{\mu\nu}(x)&&=-M_N S\cdot n {\rm\ Im\ }
{\rm Tr}(S^{\mu\nu}(x)\not\!\bar n\gamma_5)\nonumber\\
&&=i\epsilon^{\mu\nu\alpha\beta}n_\alpha \bar n_\beta M_N 
{S\cdot n\over P\cdot n} \delta (x-x_B),\nonumber\\
\sigma_b^{\mu\nu}(x)&&={P\cdot n\over 2}{\rm\ Im\ }
{\rm Tr}(iS_{\mu\nu}(x)\sigma_{\alpha\beta}\gamma_5)\bar n^\alpha S_\bot^\beta
\nonumber\\
&&=i\epsilon^{\mu\nu\alpha\beta} q_\alpha S_{\bot\beta}{2m_q\over x}
\delta (x-x_B),\nonumber\\
\sigma_{c_1}^{\mu\nu}(x_1,x)&&={M_N\over 2}P\cdot n {\rm\ Im\ }
{\rm Tr}(iS_{\mu\nu\alpha'}(x_1,x)
\not\!\bar n) \epsilon_\bot^{\alpha\delta}S^\bot_\delta
\omega^{\alpha'}_{\ \alpha}\nonumber\\
&&=i\epsilon^{\mu\nu\alpha\beta} q_\alpha S_{\bot\beta}{M_N\over 2P\cdot q}
\Big( {1\over x_1}\delta(x_1-x_B) -{1\over x}\delta(x-x_B)\Bigl),\nonumber\\
\sigma_{c_2}^{\mu\nu}(x_1,x)&&=-{M_N\over 2}P\cdot n {\rm\ Im\ }
{\rm Tr}(S_{\mu\nu\alpha'}(x_1,x) \not\!\bar n\gamma_5) S_\bot^\alpha
\omega^{\alpha'}_{\ \alpha}\nonumber\\
&&=i\epsilon^{\mu\nu\alpha\beta} q_\alpha S_{\bot\beta}{M_N\over 2P\cdot q}
\Big( {1\over x_1}\delta(x_1-x_B) +{1\over x}\delta(x-x_B)\Bigl),
\end{eqnarray}
where
\begin{eqnarray}
{\rm Im\ }S_{\mu\nu}=&&{1\over \pi}
\gamma_\mu(xp\not\! \bar n+\not\! q+m_q)
\gamma_\nu(1+{m_q\not\! n\over 2xp})\pi\delta((xp+q)^2-m_q^2)
+{\rm mirror\ diagram} ,\nonumber\\
{\rm Im\ }S_{\mu\nu\alpha'}=&&{1\over\pi}\gamma_\mu(x p\not\! \bar n+\not\! q)
(i\gamma_{\alpha'})
(x_1p\not\! \bar n+\not\! q)\gamma_\nu{i\over (x p+q)^2}
\pi\delta((x_1p+q)^2-m_q^2)\nonumber\\
&&+i\gamma_{\alpha'}{i\not\! n\over 2x_1p}
\gamma_\mu(x_1p\not\!\bar n+\not\! q)
\gamma_\nu \delta((x_1p+q)^2-m_q^2)+ {\rm mirror\ diagrams}.
\end{eqnarray}
It is easy to explicitly check that $S_{\mu\nu}$, $S_{\mu\nu\alpha'}$ and
$\sigma^{\mu\nu}$s are
separately bare electromagnetic gauge invariant in the presence of the
special propagator. This is a tremendous simplification comparing with
the conventional method employed in Refs.~\cite{et,ael}.

     After a lengthy but straightforward calculation, the antisymmetric
hadronic tensor for polarized DIS can be recast into
\begin{eqnarray}
\frac{W_{\mu\nu}^A}{2M_N}
= i\epsilon_{\mu\nu\alpha\beta}
\Bigl(n^\alpha \bar n^{\beta}\frac{S\cdot n}{P\cdot n}
g_1(x_{\text{B}})+
q^\alpha S_\bot^{\beta}
\frac {g_T(x_{\text{B}})}{P\cdot q}\Bigl),
\label{complete}
\end{eqnarray}
and the transverse polarized structure function is simply
\begin{eqnarray}
g_T(x_1)&=&g_1(x_1)+g_2(x_1)\nonumber\\
&=&\frac{1}{4x_1}\int dx_2 \Bigl[G(x_1,x_2)-G(x_2,x_1)+
\widetilde{G}(x_1,x_2)+ \widetilde{G}(x_2,x_1)+\frac{2m_q}{M_N}h_1(x_1)\delta(x_2-x_1)
\Bigr].
\label{eq:gt1}
\end{eqnarray}
Another form of $\bar g_T(x)$ in the literature~\cite{jaffe,jj} derived
from light-front QCD is
\begin{eqnarray}
g_T(x)={1\over 4M_N}\int{d\eta\over 2\pi}e^{i\eta x}
\langle PS_\bot|\bar\psi(0)\gamma_\bot \gamma_5\psi({\eta n\over P\cdot n})
|PS_\bot\rangle,
\label{eq:gt2}
\end{eqnarray}
and it looks quite different from Eq.~(\ref{eq:gt2}). In the following, we
shall briefly demonstrate that Eqs.~(\ref{eq:gt1}) and~(\ref{eq:gt2}) are
actually equivalent for the completeness of this paper.

We first observe that
\begin{eqnarray}
-d^{\alpha\beta}\gamma_5+i\epsilon_\perp^{\alpha\beta}&&=
\gamma_5(-S_{\alpha\lambda\beta\sigma}-i\epsilon_{\alpha\lambda\beta\sigma}
\gamma_5)\bar n^\lambda n^\sigma\nonumber\\
&&\equiv -\gamma_5\Sigma_{\alpha\lambda_\beta\sigma}\bar n^\lambda n^\sigma,
\end{eqnarray}
where
\begin{eqnarray}
S_{\alpha\lambda\beta\sigma}=g_{\alpha\lambda}g_{\beta\sigma}-
g_{\alpha\beta}g_{\lambda\sigma}+
g_{\alpha\sigma}g_{\lambda\beta}.
\end{eqnarray}
Using the identity $D_\alpha ={1\over 2} (\gamma_\alpha\not\!\! D+
\not\!\! D\gamma_\alpha)$, we arrive at
\begin{eqnarray}
&&S_\bot^\beta \int dx (\tilde G(x,x_1)+G(x,x_1))\nonumber\\
&&=-{1\over 2M_N (P\cdot n)}
\int {d\lambda\over 2\pi} e^{i\lambda x_1}
\langle PS|\bar\psi(0)\not n\gamma_5\Sigma_{\alpha\lambda\beta\sigma}
\bar n^\lambda n^\sigma {1\over 2}(\gamma_\alpha \not\!\! D_\bot +
\not\!\! D_\bot \gamma_\alpha)\psi({\lambda n\over P\cdot n})|PS\rangle
\nonumber\\
&&={1\over 2M_N (P\cdot n)}\int {d\lambda\over 2\pi} e^{i\lambda x_1}
\langle PS|\bar \psi_+(0) \not\!\! D_\bot \not n\gamma_5 \gamma_{\bot\beta}
\psi_+({\lambda n\over P\cdot n})|PS\rangle
\label{eq:gg}
\end{eqnarray}
where we have used
$P_+\psi=\psi_+$, $\psi^{\dag}_+ \gamma_0=\bar\psi P_-=\bar\psi_+$,
$P_+={1\over 2}\not\!\bar n \not\! n$ and $P_-={1\over 2}\not\! n\not\!\bar n$
to project out the ``good" and "bad" light-cone components of the Dirac spinor.
Applying
\begin{eqnarray}
\bar\psi_-=\psi^{\dag}_-\gamma_0=-\frac{1}{2xp}
\int {d\eta'\over 2\pi} e^{ix\eta'} \bar\psi_+
(\not\!\!\!\overleftarrow D_\bot-m_q) \not n
\end{eqnarray}
to Eq.~(\ref{eq:gg}), it is easy to obtain
\begin{eqnarray}
&&\frac{1}{4x}\int dx_1 \Bigl[\tilde G(x,x_1)+
G(x,x_1)+\frac{m_q}{M_N}h_1(x_1)\delta(x-x_1)\Bigr]\nonumber\\
&&=-{1\over 4M_N}\int {d\lambda\over 2\pi} e^{i\lambda x}
\langle PS_\bot|\bar \psi_-(0) \not\! S_\bot \gamma_5
\psi_+({\lambda n\over P\cdot n})|PS_\bot\rangle
\label{eq:ggh1}
\end{eqnarray}
and
\begin{eqnarray}
&&\frac{1}{4x}\int dx_1 \Bigl[\widetilde G(x_1,x)-
G(x_1,x)+\frac{m_q}{M_N}h_1(x_1)\delta(x-x_1)\Bigr]\nonumber\\
&&=-{1\over 4M_N}\int {d\lambda\over 2\pi} e^{i\lambda x}
\langle PS_\bot|\bar \psi_+(0) \not\! S_\bot \gamma_5
\psi_-({\lambda n\over P\cdot n})|PS_\bot\rangle.
\label{eq:ggh2}
\end{eqnarray}
With Eqs.~(\ref{eq:ggh1}) and ~({\ref{eq:ggh2}) at hand, it is obvious
that $g_T$ is in fact a measurement of the overlap between opposite
chirality partons as advertised at the beginning. The opposite chirality 
partons become two independent species without invoking the quark mass as 
the chiral symmetry breaking source and are invisible in the chirally 
invariant probe in DIS. This fact is also reflected in $\widetilde G$ and 
$G$ explicitly in the identification of the relevant matrix elements. 
Another important message in the above
derivation is the disappearance of the matrix elements with gluonic
field strength. These matrix elements bear a simple interpretation of a
charge particle scattering off a spinning top's magnetic field due to the
Lorentz force~\cite{qs}. The absence of this matrix element in $g_T$ makes 
$g_T$ a pure measurement of chiral symmetry breaking effects inside the 
nucleon. 
To summarize, we have clarified the role of the quark mass in 
defining the chiral odd transversity contribution $h_1(x)$ 
in $g_T$ in the improved parton model in DIS. 
We have also shown that the special propagator can be readily generalized 
to include the quark mass.  This technique is employed to extract the 
hidden short distant contribution buried inside the soft part. The
importance of the special propagator approach is to enable one to truly
factorize the hard and soft part and obtain an explicit gauge invariant
result twist by twist. This fact is important to ensure a parton model 
interpretation for the matrix elements thus obtained. This is because 
only a fixed number of partons are involved in the matrix elements for 
a definite twist. Therefore $h_1(x)$ in the parton model language is a 
measurement of the transversely polarized quark and anti-quark distributions
 inside the nucleon in the transverse basis, or, a measurement of chiral 
symmetry breaking effects inside the nucleon in the helicity basis. 
However, it is important to note that transversity is of 
higher twist in nature in DIS but can become a leading twist effect in other 
high energy process, say, Drell-Yan in particular. It mixes with other twist
three contributions and cannot be separately measured in DIS.

\vspace{1cm}
\acknowledgments
One of us (H.L.Y.) would like to thank J. Qiu for mentioning this problem to us.
We are grateful to P. G. Ratcliffe for useful comments.
This work was support in part by the National Science
Council of R.O.C. under Contract No. NSC87-2112-M-001-048 and
No. NSC87-2112-M-001-003.

\end{document}